\documentclass[letterpaper,twocolumn,english,prl,aps,superbib,tightenlines,floatfix,10pt]{revtex4-1}
\usepackage[T1]{fontenc}
\usepackage[latin9]{inputenc}
\usepackage{amsmath}
\usepackage{graphicx}
\usepackage{amssymb}

\usepackage{babel}

\begin{document}
\title{Ionic Coulomb Blockade in Nanopores}
\author{Matt Krems}
\author{Massimiliano Di Ventra}
\affiliation{Department of Physics, University of California, San Diego, La Jolla, CA 92093}
\date{\today}

\begin{abstract}
Understanding the dynamics of ions in nanopores is essential for applications ranging from single-molecule detection to DNA sequencing. We show both analytically and by means of molecular dynamics simulations that under specific conditions ion-ion interactions in nanopores lead to the phenomenon of ionic Coulomb blockade, namely the build-up of ions inside a nanopore with specific capacitance impeding the flow of additional ions due to Coulomb repulsion. This is the counterpart of electronic Coulomb blockade observed in mesoscopic systems. We discuss the analogies and differences with the electronic case as well as experimental situations in which this phenomenon could be detected.
\end{abstract}

\maketitle

Recently, there has been a surge of interest in nanopores due to their potential in several technological applications, the most notable being DNA sequencing and detection~\cite{zwolak08,branton08, schneider10, merchant10, garaj10, nelson10, min11}. In addition, nanopores offer unprecedented opportunities to study several fundamental physical processes related to ionic transport in confined geometries. For example, due to the tightly bound hydration layers surrounding an ion, it may be possible to observe steps in the ionic conductance due to the shedding of water molecules in the individual layers as the ion-water ``quasi-particle'' passes through the pore opening~\cite{zwolak09, zwolak10}.
This can be viewed as the classical analog of the electron conductance quantization in nanoscopic/mesoscopic systems (see, e.g., Ref.~\cite{vanWees88, wharam88}).
\begin{figure} [b]
\includegraphics[width=8.5cm]{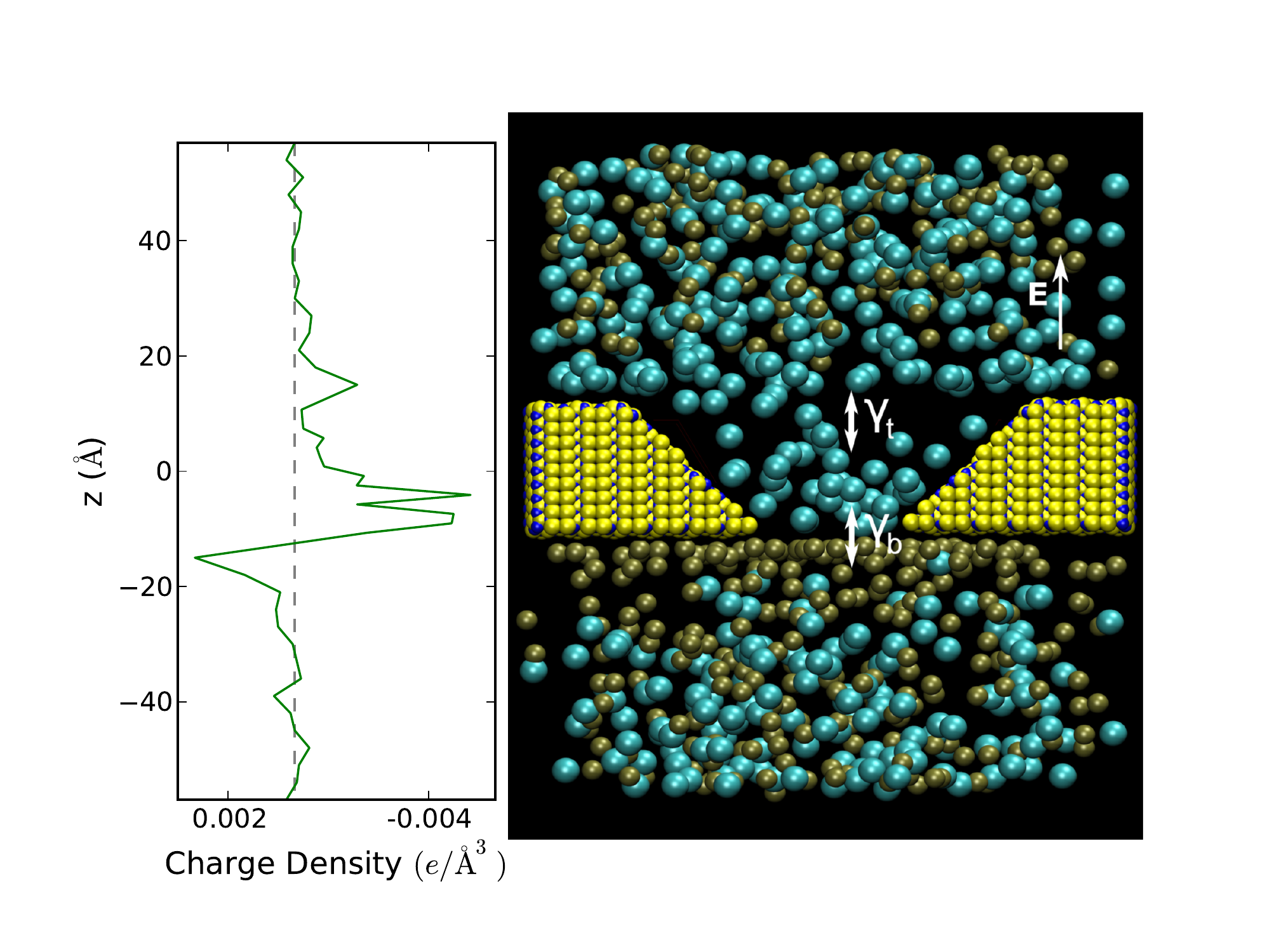}
\caption{(Color online) Right panel: a snapshot of a molecular dynamics simulation of 1M KCl solution subject to an electric field and translocating through a cone-shaped Si$_{3}$N$_{4}$ pore (indicated by the yellow and blue atoms) at a pore slope angle of 45$^{\circ}$ with a bottom opening of  7~\AA~radius and thickness of 25~\AA. Cl$^-$ ions and K$^+$ ions are colored aqua and brown, respectively. The symbols $\gamma_t$ and $\gamma_b$ indicate the rates of ion transfer at the bottom and top openings, respectively. The field
used to generate this figure is 5.0 kcal$/$(mol\AA e) which allows an easier visualization of the ions build-up. Left panel: the net charge density corresponding to the right panel configuration. It is clear that
in this case there is an accumulation of Cl$^-$ ions inside the pore while the K$^+$ ions are located mostly outside the pore.
\label{fig:densityschematic}}
\end{figure}

In this letter, we discuss another fundamental physical process which may occur in nanopores due to screened ion-ion interactions. This many-body effect originates when ions flow under the effect of an electric (or pressure) field and build up in nanopores of specific capacitances, so
that the flow of additional ions is hindered. To be specific, Fig.~\ref{fig:densityschematic} shows a typical pore geometry we have in mind with an opening wider than the other so that ions of one polarity can easily accumulate
inside the pore, while ions of opposite polarity will accumulate outside the narrow opening of the pore. The effect
we consider would then occur when the kinetic energy of an ion in solution, given approximately by the sum of the thermal energy and $M(\mu E)^2/2$ (where $\mu$ is the ion mobility, $M$ its mass and $E$ the electric field) is smaller than the capacitive energy $Q^2/2C$, with $Q$
the average number of ions in the pore, and $C$ the capacitance of the pore.

This effect is similar to the Coulomb blockade phenomenon experienced by electrons transporting through a quantum dot weakly coupled to two electrodes - in the sense that the tunneling resistance between the electrodes and the dot is much larger than the quantum of resistance~\cite{averin86}. Like the quantum case,
we also expect ionic Coulomb blockade to occur when the nanopore is ``weakly coupled'' to at least one ion reservoir. In this case, this means that the average time it takes an ion to ``cross'' the weak contact link is much longer than the
average time it takes the same ion to propagate unhindered by the pore. In other words, the ``contact'' ionic resistance of the pore with at least one reservoir has to be much larger than the resistance offered to ion flow without  the pore. From previous work~\cite{zwolak09, zwolak10} we know that such a situation may be realized when the pore opening is in the ``quantum'' regime, namely when the pore radius is small enough that an ion has to shed part
of its hydration layers to cross the pore aperture.

The analogy with the electron case therefore seems complete safe for a notable exception. In the ionic case, the Pauli exclusion principle need not be satisfied:
we can fit as many ions inside the pore as its capacitance allows without limitations from Fermi statistics. We thus expect ionic Coulomb blockade to occur as a function of the ionic concentration - which could be
thought of as controlled by a ``gate voltage'' - but not as a function of bias, namely we expect the current to show a strongly non-linear behavior as a function of molarity - within the limits of ion precipitation - at fixed
voltage, and essentially linear behavior with bias - for values below electrolysis - at fixed molarity~\footnote{Some non-linearities with bias may arise because of the different energies of the various hydration layers
(see Refs.~\cite{zwolak09, zwolak10}).}. In addition, nanopores as those shown in Fig.~\ref{fig:densityschematic} can be easily fabricated~\cite{zwolak08}. We thus expect our
predictions to be readily accessible experimentally.

Note also that the phenomenon we consider here is fundamentally different from other nonlinear ionic transport effects that have been
discussed previously in the literature~\cite{wright07,powell07, zwolak09, kuyucak94, cruzchu09}. For instance, ions in solution can form an ``ionic atmosphere'', namely
a region around a given ion in which ions of opposite charge are attracted electrostatically~\cite{wright07}. However, the ionic atmosphere is a dynamic phenomenon with typical lifetimes on the order
of $10^{-8}$ s and it can be easily destroyed at field strengths of $10^4$ V/cm~\cite{wright07}. Therefore, for the fields and molarities we consider in this work the ionic atmosphere is not relevant~\cite{wright07}.
Another deviation from Ohmic behavior has been observed experimentally in Ref.~\cite{powell07}. In this work, the addition of a small amount of divalent cations to an ionic solution leads to current oscillations and negative-incremental resistance. The phenomenon has been explained as due to the transient formation and breakup of nano-precipitates~\cite{powell07}. The nano-precipitates temporarily block the ionic current through the pore thus causing current oscillations. Again, this effect does not pertain to the present work.

Let us then start by presenting first a simple analytical model that captures the main physics of ionic Coulomb blockade. We will later corroborate these results with all-atom classical molecular dynamics simulations.
Like the electronic Coulomb blockade case (see, e.g.,~\cite{bonet02}), we can model the dynamics of ionic conduction using a rate equation approach by calculating the transition probabilities for the pore to accommodate a
certain amount of charges in addition to the background charge. Referring again to Fig.~\ref{fig:densityschematic} we consider the rates of conduction in and out of the wide opening of the pore (``top opening'') and the neck of the pore (``bottom opening''), as indicated by $\gamma_t$ and $\gamma_b$, respectively. For clarity, we refer to the ions with the charge sign that favors accumulation inside the pore. In the case of Fig.~\ref{fig:densityschematic} these
are Cl$^-$ ions, as also demonstrated by the density profile on the left panel of the same figure. The
opposite-charge ions accumulate on the outskirts of the pore entrance (see Fig.~\ref{fig:densityschematic}) and the
current blockade occurs when ions from the reservoir adjacent to the neck of the pore attempt to enter it.
The considerations we make below would then be similar, but clearly with different values of the parameters and
the direction of ion motion. If we reversed the bias direction we would obtain similar results but with the opposite charge sign accumulation inside and outside the
pore.

We assume that the rate through the bottom opening is related to the top one by
\begin{equation}
 \gamma_b = \alpha \gamma_t \;\;\;\;\;\;\;\;\;\; 0 < \alpha < 1\,,\label{alpha}
\end{equation}
with
\begin{equation}
 \gamma_t = A_t \mu n_o E\,,
 \label{eq:gamma_t}
\end{equation}
where $A_t$ is the top area of the pore, $\mu$ is the ionic mobility, $n_0$ is the ionic density, and $E=V/d$ is the electric field, where $V$ is the voltage and $d$ is the length of the pore~\cite{diventra08}. The parameter $\alpha$ takes into account the difference in the two rates due to the partial shedding of the hydration layers around the ions when they cross the bottom neck, and the fact that the radius of the pore opening is smaller at the neck than at the top.

In order to be able to solve the rate equations analytically we assume only two possible ``ionic states''. The first state with transition probability $P_0$ corresponds to only the background charge in the pore, namely to $\lfloor{n_0\Omega_p}\rfloor$ ions, where $n_0$ is the background density, $\Omega_p$ is the volume of the pore, and the symbol $\lfloor \cdots \rfloor$ represents the floor function which will make the product $n_0\Omega_p$ an integer.
The second state with probability $P_1$ is that of the background charge plus one single extra ion, namely $\lfloor{n_0\Omega_p}\rfloor + 1$ ions in the pore. While this may seem like an oversimplified situation, it captures the main trends observed with the
molecular dynamics simulations as we show below. The Markovian equation for the transition probability $P_0$ is then
\begin{equation}
 \frac{dP_0}{dt} = \Gamma_{1\rightarrow 0}P_1 - \Gamma_{0\rightarrow 1}P_0,
 \label{eq:rate}
\end{equation}
where $\Gamma_{0\rightarrow 1}$ is the transition rate of going from state $0$ to state $1$, and  $\Gamma_{1\rightarrow 0}$ is the reverse process. The equation of motion for $P_1$ is easily obtained by interchanging 0 with 1 in Eq.~\ref{eq:rate}.

\begin{figure} [b]
\centering
\includegraphics[width=7.5cm]{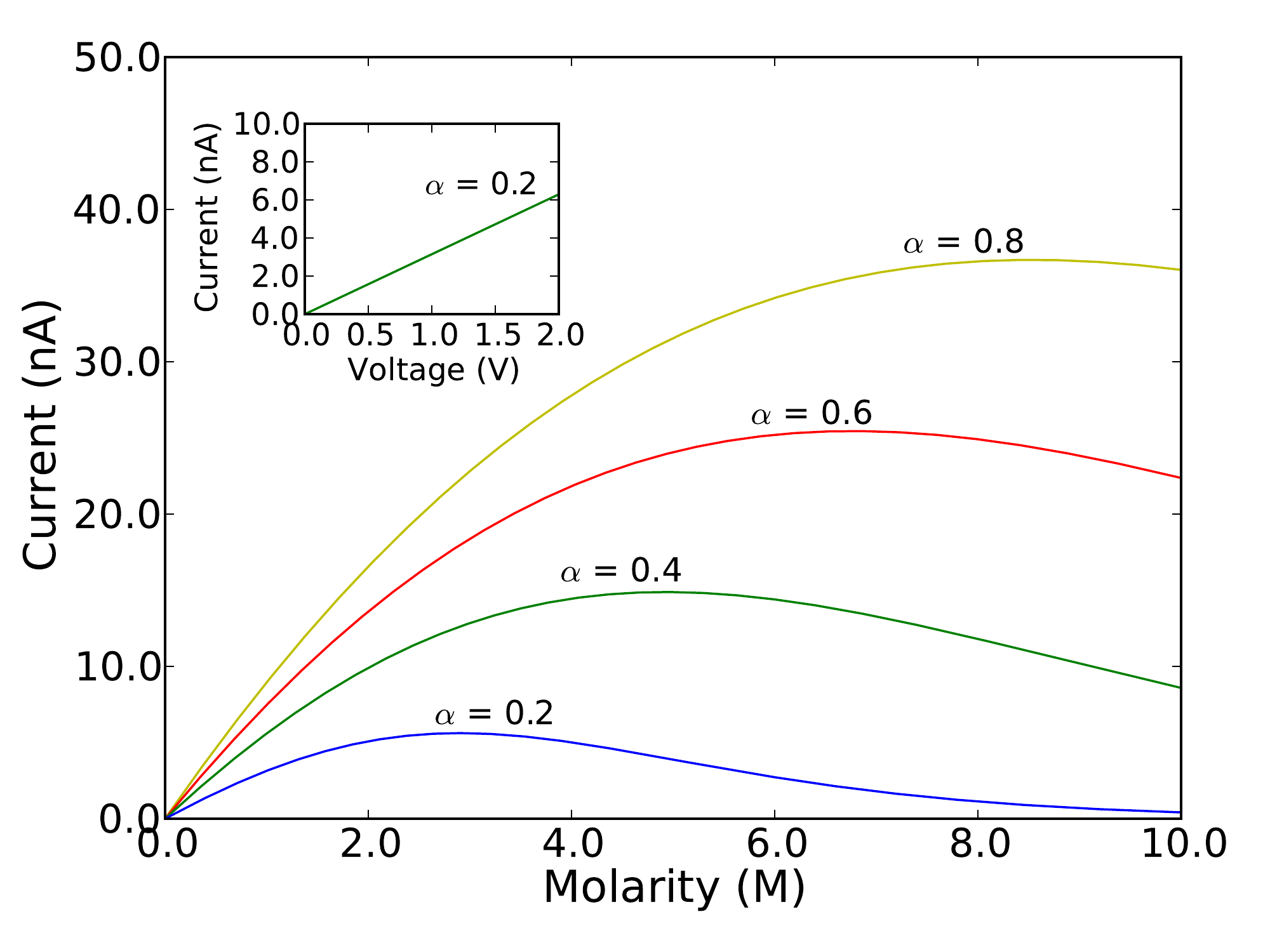}
\caption{(Color online) Current from Eq.~(\ref{finalcurr}) as a function of the ion concentration at a fixed voltage of $V = 1.0$~V for various pore neck openings as represented by the parameter $\alpha$ (larger $\alpha$ implies
larger neck opening). The inset shows the current, at a concentration of 1.0 M, as a function of bias for $\alpha=0.2$. See text for all other parameters.
\label{fig:model}}
\end{figure}

The transition rate $\Gamma_{0\rightarrow 1}$ contains the capacitive barrier experienced by an ion entering the pore already
occupied by $\lfloor{n_0\Omega_p}\rfloor$ ions. Using the steady-state Nernst-Planck equation it can then be written as~\cite{zwolak10}
\begin{equation}
 \Gamma_{0 \rightarrow 1} = \gamma_t \exp{(-\epsilon_C/kT)},
\end{equation}
where $\epsilon_C$ is a single-particle capacitive energy barrier, $k$ the Boltzmann constant, and $T$ the temperature. We can write the single-particle capacitive energy as~\cite{bonet02}
\begin{equation}
 \epsilon_C=e^2\left(\frac{2 \lfloor n_0\Omega_p \rfloor+1}{2C}\right)
 \label{eq:e_C}
\end{equation}
where $e$ is the elementary charge and $C$ is the capacitance of the pore. Since the capacitance of the pore is linearly related to the ratio of the surface areas of the top and bottom openings, it must vary according to the
parameter $\alpha$ we have introduced in Eq.~(\ref{alpha}). We then assume $C = \alpha C_0$ with $C_0$ some reasonable experimental value for the capacitance. The rate $\Gamma_{1 \rightarrow 0}$ is instead simply
\begin{equation}
 \Gamma_{1 \rightarrow 0} = \gamma_b,
\end{equation}
because there is no capacitive barrier for this process. We also note that, in principle, there is a finite rate for an ion to move in the direction opposite to that determined by the electric field. This rate would contribute
to both $\Gamma_{0 \rightarrow 1}$ and $\Gamma_{1 \rightarrow 0}$. However, this
process is exponentially suppressed for the biases we consider here, and we can thus safely ignore it \footnote{There is also an entropic barrier associated with ions entering the pore, but this is also negligible compared to the capacitive barrier.}.

At steady state, $dP/dt=0$, and from Eq.~(\ref{eq:rate}) - and the equivalent equation for $P_1$ - we obtain the steady-state probabilities
\begin{equation}
 P_0 = \frac{\Gamma_{1 \rightarrow 0}}{\Gamma_{0 \rightarrow 1} + \Gamma_{1 \rightarrow 0}} =\ \frac{\gamma_b}{\gamma_t\exp{(-\epsilon_C/kT)}+\gamma_b}
\end{equation}
\begin{equation}
 P_1 = \frac{\Gamma_{0 \rightarrow 1}}{\Gamma_{0 \rightarrow 1} + \Gamma_{1 \rightarrow 0}} = \frac{\gamma_t\exp{(-\epsilon_C/kT)}}{\gamma_t\exp{(-\epsilon_C/kT)}+\gamma_b}
\end{equation}

Finally, the current at steady state is the same everywhere and we evaluate it across the neck of the pore
\begin{equation}
 I_b = e \left(\Gamma_{1\rightarrow 0}^b P_1 - \Gamma_{0\rightarrow 1}^bP_0\right),
\end{equation}
where the superscript $b$  indicates that we retain only the terms corresponding to the bottom part of the transition rates. In the present case, this gives
\begin{equation}
 I_b = e \alpha \gamma_t \left(\frac{\exp{(-\epsilon_C/kT)}}{\exp{(-\epsilon_C/kT)}+\alpha}\right), \label{finalcurr}
\end{equation}
with $\gamma_t$ and $\epsilon_C$ given by Eqs.~(\ref{eq:gamma_t}) and~(\ref{eq:e_C}), respectively. This equation shows indeed what we were expecting: the current is linearly dependent on bias (via the parameter $\gamma_t$)
for a fixed ion concentration, and it saturates - and eventually decreases - with increasing concentration according to the difference between the top and bottom openings (parameter $\alpha$). This is shown in Fig.~\ref{fig:model}
where we have used the parameters $A_t = 7.8$ nm$^3$, $d=25$~\AA, $\mu = 7\times10^{-8}$~m$^2$V$^{-1}$s$^{-1}$,  $T=295$ K, and $C/\alpha=1.0$~fF. The latter is a reasonable value for a nanopore of our size~\cite{krems10}. With these parameters we indeed find that the typical drift contribution to the kinetic energy of an ion in solution is some fraction of 1 meV,
which (as the thermal energy) is smaller than the capacitive barrier (on the order of several tens of meVs).

We now show that this model captures the main physics of this phenomenon by performing all-atom molecular dynamics simulations using NAMD2~\cite{phillips05}. The pores are made of
25~\AA$\;$thick silicon nitride material in the $\beta$-phase and they have a conical shape with a slope angle of 45$^{\circ}$ (see Fig.~\ref{fig:densityschematic}). We vary the bottom opening from a radius $r=5$ \AA$\,$ to a radius $r=10$ \AA. We then
introduce a given concentration of KCl while keeping the temperature fixed at 295 K. By introducing an external constant electric field we can then probe the ionic conductance. Other details of the simulations are reported in Ref.~\cite{krems10}.

\begin{figure} [t]
\centering
\includegraphics[width=7.5cm]{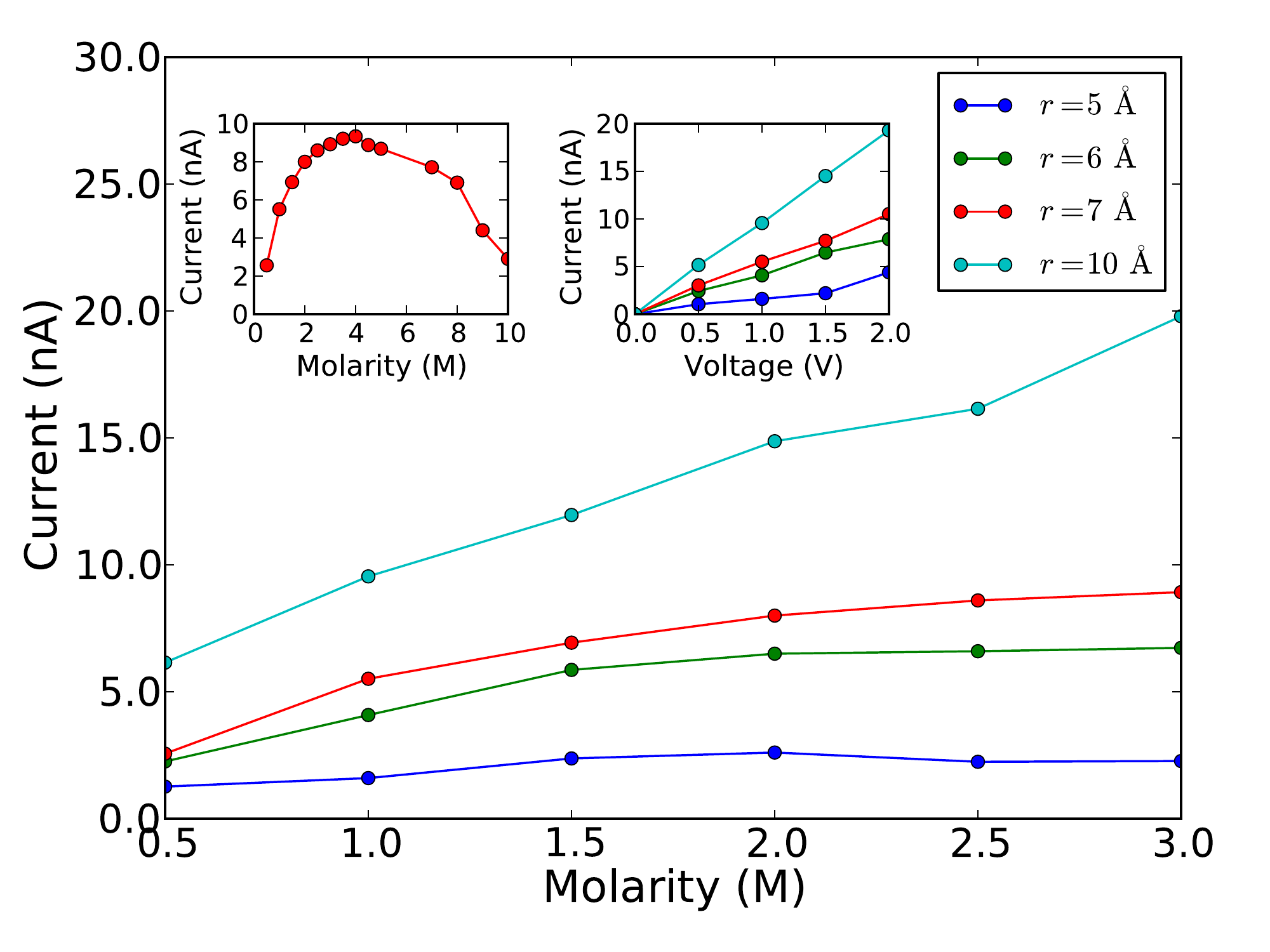}
\caption{(Color online) Current as a function of KCl molarity for various neck radii and at a fixed bias of 1.0 V. For relevant experimental molarities, we observe an almost linear increase of the current for a radius $r=10\AA$
and a non-linear behavior approaching a saturation at a given molarity for smaller radii. The current saturation and decrease is shown explicitly in the left inset for a 7~\AA~neck radius pore. We also show the increase in current with voltage in the other inset. \label{fig:current}}
\end{figure}

The results of these simulations are plotted in Fig.~\ref{fig:current}. As predicted by Eq.~(\ref{finalcurr}) our calculations show an almost linear behavior of the current as a function of bias, and for a fixed bias
a saturation of the current as a function of ion concentration which is more pronounced for pores with smaller neck radius. In the inset of Fig.~\ref{fig:current} we also
show explicitly current saturation and decrease for a 7~\AA~neck radius pore, which has to be compared with the results reported in Fig.~\ref{fig:model}. Note, however, that current saturation and decrease occur at molarities
well above the ionic precipitation limit of KCl of about 3.5 M. We thus expect these two features not to be directly visible for this configuration.

Finally, we discuss conditions under which the phenomenon of ionic Coulomb blockade may be better resolved. It is definitely true that nanopores of a V-shape as that shown in Fig.~\ref{fig:densityschematic} can be easily fabricated~\cite{zwolak08}. However, the amount of surface charges on the internal walls of the pore is not so straightforward to control. Nevertheless, surface charges are beneficial to the observation of this effect.
Indeed, if a certain amount of, say, positive charges are present at the surface of the pore walls, more Cl$^-$ ions would accumulate inside the pore even without the need of an electric field. Everything else
being equal, the predicted non-linearities as a function of molarity would then occur at lower concentrations. Better yet would be if these surface charges could be modified electrically. This has been recently achieved
experimentally by placing electrodes inside nanopores of a shape similar to the one considered in this paper~\cite{nam09,taniguchi09,ivanov11}. The predicted phenomenon is therefore within reach of these experiments. Note also that the same phenomenon would occur in the case of a straight-wall (e.g., cylindrical) pore provided the openings are small enough that the hydration layers need to be partly shed. In this case, using molecular dynamics simulations we have also observed
ionic current blockade as a function of molarity (not reported in this paper). Here, however, we expect the pore wall charges to play a much more important role than in the case of V-shaped pores.
In this respect, experiments with more well-controlled synthetic pores, e.g., nanotube pores~\cite{liu10} may be particularly suitable to verify this phenomenon. We leave all these studies for future work.

In conclusion, we have predicted that ionic Coulomb blockade effects should be observable in nanopores of specific geometries, when ions accumulate in the pore thus impeding the flow of like-charge ions. We have corroborated these predictions with all-atom molecular dynamics simulations. The ensuing non-linearities as a function of ion concentration are within reach of experimental verification. This phenomenon parallels the one observed in electronic
transport across quantum dots weakly coupled to two electrodes. This effect may play a significant role in various applications of nanopores presently pursued, and is of importance to understand ionic conduction in confined geometries.

We thank Jim Wilson and Alexander Stotland for useful discussions. Financial support from the NIH-National Human Genome Research Institute is gratefully acknowledged.

\bibliography{ionicblockade}

\begin{thebibliography}{10}%
\makeatletter
\providecommand \@ifxundefined [1]{%
 \ifx #1\undefined \expandafter \@firstoftwo
 \else \expandafter \@secondoftwo
\fi
}%
\providecommand \@ifnum [1]{%
 \ifnum #1\expandafter \@firstoftwo
 \else \expandafter \@secondoftwo
\fi
}%
\providecommand \enquote [1]{``#1''}%
\providecommand \bibnamefont  [1]{#1}%
\providecommand \bibfnamefont [1]{#1}%
\providecommand \citenamefont [1]{#1}%
\providecommand\href[0]{\@sanitize\@href}%
\providecommand\@href[1]{\endgroup\@@startlink{#1}\endgroup\@@href}%
\providecommand\@@href[1]{#1\@@endlink}%
\providecommand \@sanitize [0]{\begingroup\catcode`\&12\catcode`\#12\relax}%
\@ifxundefined \pdfoutput {\@firstoftwo}{%
 \@ifnum{\z@=\pdfoutput}{\@firstoftwo}{\@secondoftwo}%
}{%
 \providecommand\@@startlink[1]{\leavevmode\special{html:<a href="#1">}}%
 \providecommand\@@endlink[0]{\special{html:</a>}}%
}{%
 \providecommand\@@startlink[1]{%
  \leavevmode
  \pdfstartlink
   attr{/Border[0 0 1 ]/H/I/C[0 1 1]}%
   user{/Subtype/Link/A<</Type/Action/S/URI/URI(#1)>>}%
  \relax
 }%
 \providecommand\@@endlink[0]{\pdfendlink}%
}%
\providecommand \url  [0]{\begingroup\@sanitize \@url }%
\providecommand \@url [1]{\endgroup\@href {#1}{\urlprefix}}%
\providecommand \urlprefix [0]{URL }%
\providecommand \Eprint[0]{\href }%
\@ifxundefined \urlstyle {%
  \providecommand \doi [1]{doi:\discretionary{}{}{}#1}%
}{%
  \providecommand \doi [0]{doi:\discretionary{}{}{}\begingroup
  \urlstyle{rm}\Url }%
}%
\providecommand \doibase [0]{http://dx.doi.org/}%
\providecommand \Doi[1]{\href{\doibase#1}}%
\providecommand \bibAnnote [3]{%
  \BibitemShut{#1}%
  \begin{quotation}\noindent
    \textsc{Key:}\ #2\\\textsc{Annotation:}\ #3%
  \end{quotation}%
}%
\providecommand \bibAnnoteFile [2]{%
  \IfFileExists{#2}{\bibAnnote {#1} {#2} {\input{#2}}}{}%
}%
\providecommand \typeout [0]{\immediate \write \m@ne }%
\providecommand \selectlanguage [0]{\@gobble}%
\providecommand \bibinfo [0]{\@secondoftwo}%
\providecommand \bibfield [0]{\@secondoftwo}%
\providecommand \translation [1]{[#1]}%
\providecommand \BibitemOpen[0]{}%
\providecommand \bibitemStop [0]{}%
\providecommand \bibitemNoStop [0]{.\EOS\space}%
\providecommand \EOS [0]{\spacefactor3000\relax}%
\providecommand \BibitemShut [1]{\csname bibitem#1\endcsname}%
\bibitem{zwolak08}%
  \BibitemOpen
  \bibfield{author}{%
  \bibinfo {author} {\bibfnamefont{M.}~\bibnamefont{Zwolak}}\ and\ \bibinfo
  {author} {\bibfnamefont{M.}~\bibnamefont{{Di Ventra}}},\ }%
  \bibfield{journal}{%
  \bibinfo {journal} {Rev. Mod. Phys.}\ }%
  \textbf{\bibinfo {volume} {80}},\ \bibinfo {pages} {141} (\bibinfo {year}
  {2008})%
  \bibAnnoteFile{NoStop}{zwolak08}%
\bibitem{branton08}%
  \BibitemOpen
  \bibfield{author}{%
  \bibinfo {author} {\bibfnamefont{D.}~\bibnamefont{Branton}}, \bibinfo
  {author} {\bibfnamefont{D.~W.}\ \bibnamefont{Deamer}}, \bibinfo {author}
  {\bibfnamefont{A.}~\bibnamefont{Marziali}}, \bibinfo {author}
  {\bibfnamefont{H.}~\bibnamefont{Bayley}}, \bibinfo {author}
  {\bibfnamefont{S.~A.}\ \bibnamefont{Benner}}, \bibinfo {author}
  {\bibfnamefont{T.}~\bibnamefont{Butler}}, \bibinfo {author}
  {\bibfnamefont{M.}~\bibnamefont{{Di Ventra}}}, \bibinfo {author}
  {\bibfnamefont{S.}~\bibnamefont{Garaj}}, \bibinfo {author}
  {\bibfnamefont{A.}~\bibnamefont{Hibbs}}, \bibinfo {author}
  {\bibfnamefont{X.}~\bibnamefont{Huang}}, \bibinfo {author}
  {\bibfnamefont{S.~B.}\ \bibnamefont{Jovanovich}}, \bibinfo {author}
  {\bibfnamefont{P.~S.}\ \bibnamefont{Krstic}}, \bibinfo {author}
  {\bibfnamefont{S.}~\bibnamefont{Lindsay}}, \bibinfo {author}
  {\bibfnamefont{X.~S.}\ \bibnamefont{Ling}}, \bibinfo {author}
  {\bibfnamefont{C.~H.}\ \bibnamefont{Mastrangelo}}, \bibinfo {author}
  {\bibfnamefont{A.}~\bibnamefont{Meller}}, \bibinfo {author}
  {\bibfnamefont{J.~S.}\ \bibnamefont{Oliver}}, \bibinfo {author}
  {\bibfnamefont{Y.~V.}\ \bibnamefont{Pershin}}, \bibinfo {author}
  {\bibfnamefont{J.~M.}\ \bibnamefont{Ramsey}}, \bibinfo {author}
  {\bibfnamefont{R.}~\bibnamefont{Riehn}}, \bibinfo {author}
  {\bibfnamefont{G.~V.}\ \bibnamefont{Soni}}, \bibinfo {author}
  {\bibfnamefont{V.}~\bibnamefont{Tabard-Cossa}}, \bibinfo {author}
  {\bibfnamefont{M.}~\bibnamefont{Wanunu}}, \bibinfo {author}
  {\bibfnamefont{M.}~\bibnamefont{Wiggin}},\ and\ \bibinfo {author}
  {\bibfnamefont{J.~A.}\ \bibnamefont{Schloss}},\ }%
  \bibfield{journal}{%
  \bibinfo {journal} {Nat. Biotechnol.}\ }%
  \textbf{\bibinfo {volume} {26}},\ \bibinfo {pages} {1146} (\bibinfo {year}
  {2008})%
  \bibAnnoteFile{NoStop}{branton08}%
\bibitem{schneider10}%
  \BibitemOpen
  \bibfield{author}{%
  \bibinfo {author} {\bibfnamefont{G.~F.}\ \bibnamefont{Schneider}}, \bibinfo
  {author} {\bibfnamefont{S.~W.}\ \bibnamefont{Kowalczyk}}, \bibinfo {author}
  {\bibfnamefont{V.~E.}\ \bibnamefont{Calado}}, \bibinfo {author}
  {\bibfnamefont{G.}~\bibnamefont{Pandraud}}, \bibinfo {author}
  {\bibfnamefont{H.~W.}\ \bibnamefont{Zandbergen}}, \bibinfo {author}
  {\bibfnamefont{L.~M.~K.}\ \bibnamefont{Vandersypen}},\ and\ \bibinfo {author}
  {\bibfnamefont{C.}~\bibnamefont{Dekker}},\ }%
  \bibfield{journal}{%
  \bibinfo {journal} {Nano Lett.}\ }%
  \textbf{\bibinfo {volume} {10}},\ \bibinfo {pages} {3163} (\bibinfo {year}
  {2010})%
  \bibAnnoteFile{NoStop}{schneider10}%
\bibitem{merchant10}%
  \BibitemOpen
  \bibfield{author}{%
  \bibinfo {author} {\bibfnamefont{C.~A.}\ \bibnamefont{Merchant}}, \bibinfo
  {author} {\bibfnamefont{K.}~\bibnamefont{Healy}}, \bibinfo {author}
  {\bibfnamefont{M.}~\bibnamefont{Wanunu}}, \bibinfo {author}
  {\bibfnamefont{V.}~\bibnamefont{Ray}}, \bibinfo {author}
  {\bibfnamefont{N.}~\bibnamefont{Peterman}}, \bibinfo {author}
  {\bibfnamefont{J.}~\bibnamefont{Bartel}}, \bibinfo {author}
  {\bibfnamefont{M.~D.}\ \bibnamefont{Fischbein}}, \bibinfo {author}
  {\bibfnamefont{K.}~\bibnamefont{Venta}}, \bibinfo {author}
  {\bibfnamefont{Z.}~\bibnamefont{Luo}}, \bibinfo {author}
  {\bibfnamefont{A.~T.~C.}\ \bibnamefont{Johnson}},\ and\ \bibinfo {author}
  {\bibfnamefont{M.}~\bibnamefont{Drndic}},\ }%
  \bibfield{journal}{%
  \bibinfo {journal} {Nano Lett.}\ }%
  \textbf{\bibinfo {volume} {10}},\ \bibinfo {pages} {2915} (\bibinfo {year}
  {2010})%
  \bibAnnoteFile{NoStop}{merchant10}%
\bibitem{garaj10}%
  \BibitemOpen
  \bibfield{author}{%
  \bibinfo {author} {\bibfnamefont{S.}~\bibnamefont{Garaj}}, \bibinfo {author}
  {\bibfnamefont{W.}~\bibnamefont{Hubbard}}, \bibinfo {author}
  {\bibfnamefont{A.}~\bibnamefont{Reina}}, \bibinfo {author}
  {\bibfnamefont{J.}~\bibnamefont{Kong}}, \bibinfo {author}
  {\bibfnamefont{D.}~\bibnamefont{Branton}},\ and\ \bibinfo {author}
  {\bibfnamefont{J.~A.}\ \bibnamefont{Golovchenko}},\ }%
  \bibfield{journal}{%
  \bibinfo {journal} {Nature}\ }%
  \textbf{\bibinfo {volume} {467}},\ \bibinfo {pages} {190} (\bibinfo {year}
  {2010})%
  \bibAnnoteFile{NoStop}{garaj10}%
\bibitem{nelson10}%
  \BibitemOpen
  \bibfield{author}{%
  \bibinfo {author} {\bibfnamefont{B.~Z.}\ \bibnamefont{Tammie~Nelson}}\ and\
  \bibinfo {author} {\bibfnamefont{O.~V.}\ \bibnamefont{Prezhdo}},\ }%
  \bibfield{journal}{%
  \bibinfo {journal} {Nano Lett.}\ }%
  \textbf{\bibinfo {volume} {10}},\ \bibinfo {pages} {3237} (\bibinfo {year}
  {2010})%
  \bibAnnoteFile{NoStop}{nelson10}%
\bibitem{min11}%
  \BibitemOpen
  \bibfield{author}{%
  \bibinfo {author} {\bibfnamefont{S.~K.}\ \bibnamefont{Min}}, \bibinfo
  {author} {\bibfnamefont{W.~Y.}\ \bibnamefont{Kim}}, \bibinfo {author}
  {\bibfnamefont{Y.}~\bibnamefont{Cho}},\ and\ \bibinfo {author}
  {\bibfnamefont{K.~S.}\ \bibnamefont{Kim}},\ }%
  \bibfield{journal}{%
  \bibinfo {journal} {Nat. Nanotechnology}\ }%
  \textbf{\bibinfo {volume} {6}},\ \bibinfo {pages} {162} (\bibinfo {year}
  {2011})%
  \bibAnnoteFile{NoStop}{min11}%
\bibitem{zwolak09}%
  \BibitemOpen
  \bibfield{author}{%
  \bibinfo {author} {\bibfnamefont{M.}~\bibnamefont{Zwolak}}, \bibinfo {author}
  {\bibfnamefont{J.}~\bibnamefont{Lagerqvist}},\ and\ \bibinfo {author}
  {\bibfnamefont{M.}~\bibnamefont{{Di Ventra}}},\ }%
  \bibfield{journal}{%
  \bibinfo {journal} {Phys. Rev. Lett.}\ }%
  \textbf{\bibinfo {volume} {103}},\ \bibinfo {pages} {128102} (\bibinfo {year}
  {2009})%
  \bibAnnoteFile{NoStop}{zwolak09}%
\bibitem{zwolak10}%
  \BibitemOpen
  \bibfield{author}{%
  \bibinfo {author} {\bibfnamefont{M.}~\bibnamefont{Zwolak}}, \bibinfo {author}
  {\bibfnamefont{J.}~\bibnamefont{Wilson}},\ and\ \bibinfo {author}
  {\bibfnamefont{M.}~\bibnamefont{{Di Ventra}}},\ }%
  \bibfield{journal}{%
  \bibinfo {journal} {J. Phys. Condens. Matter}\ }%
  \textbf{\bibinfo {volume} {22}},\ \bibinfo {pages} {454126} (\bibinfo {year}
  {2010})%
  \bibAnnoteFile{NoStop}{zwolak10}%
\bibitem{vanWees88}%
  \BibitemOpen
  \bibfield{author}{%
  \bibinfo {author} {\bibfnamefont{B.~J.}\ \bibnamefont{{van Wees}}}, \bibinfo
  {author} {\bibfnamefont{H.}~\bibnamefont{{van Houten}}}, \bibinfo {author}
  {\bibfnamefont{C.~W.~J.}\ \bibnamefont{Beenakker}}, \bibinfo {author}
  {\bibfnamefont{J.~G.}\ \bibnamefont{Williamson}}, \bibinfo {author}
  {\bibfnamefont{L.~P.}\ \bibnamefont{Kouwenhoven}}, \bibinfo {author}
  {\bibfnamefont{D.}~\bibnamefont{{van der Marel}}},\ and\ \bibinfo {author}
  {\bibfnamefont{C.~T.}\ \bibnamefont{Foxon}},\ }%
  \bibfield{journal}{%
  \bibinfo {journal} {Phys. Rev. Lett.}}%
   (\bibinfo {year} {1988})%
  \bibAnnoteFile{NoStop}{vanWees88}%
\bibitem{wharam88}%
  \BibitemOpen
  \bibfield{author}{%
  \bibinfo {author} {\bibfnamefont{D.~A.}\ \bibnamefont{Wharam}}, \bibinfo
  {author} {\bibfnamefont{T.~J.}\ \bibnamefont{Thornton}}, \bibinfo {author}
  {\bibfnamefont{R.}~\bibnamefont{Newbury}}, \bibinfo {author}
  {\bibfnamefont{M.}~\bibnamefont{Pepper}}, \bibinfo {author}
  {\bibfnamefont{H.}~\bibnamefont{Ahmed}}, \bibinfo {author}
  {\bibfnamefont{J.~E.~F.}\ \bibnamefont{Frost}}, \bibinfo {author}
  {\bibfnamefont{D.~G.}\ \bibnamefont{Hasko}}, \bibinfo {author}
  {\bibfnamefont{D.~C.}\ \bibnamefont{Peacock}}, \bibinfo {author}
  {\bibfnamefont{D.~A.}\ \bibnamefont{Ritchie}},\ and\ \bibinfo {author}
  {\bibfnamefont{G.~A.~C.}\ \bibnamefont{Jones}},\ }%
  \bibfield{journal}{%
  \bibinfo {journal} {J. Phys. C}\ }%
  \textbf{\bibinfo {volume} {21}},\ \bibinfo {pages} {L209} (\bibinfo {year}
  {1988})%
  \bibAnnoteFile{NoStop}{wharam88}%
\bibitem{averin86}%
  \BibitemOpen
  \bibfield{author}{%
  \bibinfo {author} {\bibfnamefont{D.}~\bibnamefont{Averin}}\ and\ \bibinfo
  {author} {\bibfnamefont{K.}~\bibnamefont{Likharev}},\ }%
  \bibfield{journal}{%
  \bibinfo {journal} {J. Low Temp. Phys.}\ }%
  \textbf{\bibinfo {volume} {62}},\ \bibinfo {pages} {345} (\bibinfo {year}
  {1986})%
  \bibAnnoteFile{NoStop}{averin86}%
\bibitem{Note1}%
  \BibitemOpen
  \bibinfo {note} {Some non-linearities with bias may arise because of the
  different energies of the various hydration layers (see Refs.~\cite
  {zwolak09, zwolak10}).}%
  \bibAnnoteFile{Stop}{Note1}%
\bibitem{wright07}%
  \BibitemOpen
  \bibfield{author}{%
  \bibinfo {author} {\bibfnamefont{M.~R.}\ \bibnamefont{Wright}},\ }%
  \emph{\bibinfo {title} {An Introduction to Aqueous Electrolyte Solutions}}\
  (\bibinfo {publisher} {Wiley},\ \bibinfo {year} {2007})%
  \bibAnnoteFile{NoStop}{wright07}%
\bibitem{powell07}%
  \BibitemOpen
  \bibfield{author}{%
  \bibinfo {author} {\bibfnamefont{M.~R.}\ \bibnamefont{Powell}}, \bibinfo
  {author} {\bibfnamefont{M.}~\bibnamefont{Sullivan}}, \bibinfo {author}
  {\bibfnamefont{I.}~\bibnamefont{Vlassiouk}}, \bibinfo {author}
  {\bibfnamefont{D.}~\bibnamefont{Constantin}}, \bibinfo {author}
  {\bibfnamefont{O.}~\bibnamefont{Sudre}}, \bibinfo {author}
  {\bibfnamefont{C.~C.}\ \bibnamefont{Martens}}, \bibinfo {author}
  {\bibfnamefont{R.~S.}\ \bibnamefont{Eisenberg}},\ and\ \bibinfo {author}
  {\bibfnamefont{Z.~S.}\ \bibnamefont{Siwy}},\ }%
  \bibfield{journal}{%
  \bibinfo {journal} {Nat. Nanotechnology}\ }%
  \textbf{\bibinfo {volume} {3}},\ \bibinfo {pages} {51} (\bibinfo {year}
  {2007})%
  \bibAnnoteFile{NoStop}{powell07}%
\bibitem{kuyucak94}%
  \BibitemOpen
  \bibfield{author}{%
  \bibinfo {author} {\bibfnamefont{S.}~\bibnamefont{Kuyucak}}\ and\ \bibinfo
  {author} {\bibfnamefont{S.-H.}\ \bibnamefont{Chung}},\ }%
  \bibfield{journal}{%
  \bibinfo {journal} {Biophys. Chem.}\ }%
  \textbf{\bibinfo {volume} {52}},\ \bibinfo {pages} {15} (\bibinfo {year}
  {1994})%
  \bibAnnoteFile{NoStop}{kuyucak94}%
\bibitem{cruzchu09}%
  \BibitemOpen
  \bibfield{author}{%
  \bibinfo {author} {\bibfnamefont{E.~R.}\ \bibnamefont{Cruz-Chu}}, \bibinfo
  {author} {\bibfnamefont{A.}~\bibnamefont{Aksimentiev}},\ and\ \bibinfo
  {author} {\bibfnamefont{K.}~\bibnamefont{Schulten}},\ }%
  \bibfield{journal}{%
  \bibinfo {journal} {J. Phys. Chem.}\ }%
  \textbf{\bibinfo {volume} {113}},\ \bibinfo {pages} {1850} (\bibinfo {year}
  {2009})%
  \bibAnnoteFile{NoStop}{cruzchu09}%
\bibitem{bonet02}%
  \BibitemOpen
  \bibfield{author}{%
  \bibinfo {author} {\bibfnamefont{E.}~\bibnamefont{Bonet}}, \bibinfo {author}
  {\bibfnamefont{M.~M.}\ \bibnamefont{Deshmukh}},\ and\ \bibinfo {author}
  {\bibfnamefont{D.}~\bibnamefont{Ralph}},\ }%
  \bibfield{journal}{%
  \bibinfo {journal} {Phys. Rev. B}\ }%
  \textbf{\bibinfo {volume} {65}} (\bibinfo {year} {2002})%
  \bibAnnoteFile{NoStop}{bonet02}%
\bibitem{diventra08}%
  \BibitemOpen
  \bibfield{author}{%
  \bibinfo {author} {\bibfnamefont{M.}~\bibnamefont{{Di Ventra}}},\ }%
  \emph{\bibinfo {title} {Electrical Transport in Nanoscale Systems}}\
  (\bibinfo {publisher} {Cambridge University Press},\ \bibinfo {year} {2008})%
  \bibAnnoteFile{NoStop}{diventra08}%
\bibitem{Note2}%
  \BibitemOpen
  \bibinfo {note} {There is also an entropic barrier associated with ions
  entering the pore, but this is also negligible compared to the capacitive
  barrier.}%
  \bibAnnoteFile{Stop}{Note2}%
\bibitem{krems10}%
  \BibitemOpen
  \bibfield{author}{%
  \bibinfo {author} {\bibfnamefont{M.}~\bibnamefont{Krems}}, \bibinfo {author}
  {\bibfnamefont{Y.~V.}\ \bibnamefont{Pershin}},\ and\ \bibinfo {author}
  {\bibfnamefont{M.}~\bibnamefont{{Di Ventra}}},\ }%
  \bibfield{journal}{%
  \bibinfo {journal} {Nano Lett.}\ }%
  \textbf{\bibinfo {volume} {10}},\ \bibinfo {pages} {2674} (\bibinfo {year}
  {2010})%
  \bibAnnoteFile{NoStop}{krems10}%
\bibitem{phillips05}%
  \BibitemOpen
  \bibfield{author}{%
  \bibinfo {author} {\bibfnamefont{J.~C.}\ \bibnamefont{Phillips}}, \bibinfo
  {author} {\bibfnamefont{R.}~\bibnamefont{Braun}}, \bibinfo {author}
  {\bibfnamefont{W.}~\bibnamefont{Wand}}, \bibinfo {author}
  {\bibfnamefont{J.}~\bibnamefont{Gumbart}}, \bibinfo {author}
  {\bibfnamefont{E.}~\bibnamefont{Tajkhorshid}}, \bibinfo {author}
  {\bibfnamefont{E.}~\bibnamefont{Villa}}, \bibinfo {author}
  {\bibfnamefont{C.}~\bibnamefont{Chipot}}, \bibinfo {author}
  {\bibfnamefont{R.~D.}\ \bibnamefont{Skeel}}, \bibinfo {author}
  {\bibfnamefont{L.}~\bibnamefont{Kale}},\ and\ \bibinfo {author}
  {\bibfnamefont{K.}~\bibnamefont{Schulten}},\ }%
  \bibfield{journal}{%
  \bibinfo {journal} {J. Comp. Chem.}\ }%
  \textbf{\bibinfo {volume} {26}},\ \bibinfo {pages} {1781} (\bibinfo {year}
  {2005})%
  \bibAnnoteFile{NoStop}{phillips05}%
\bibitem{nam09}%
  \BibitemOpen
  \bibfield{author}{%
  \bibinfo {author} {\bibfnamefont{S.-W.}\ \bibnamefont{Nam}}, \bibinfo
  {author} {\bibfnamefont{M.~J.}\ \bibnamefont{Rooks}}, \bibinfo {author}
  {\bibfnamefont{K.-B.}\ \bibnamefont{Kim}},\ and\ \bibinfo {author}
  {\bibfnamefont{S.~M.}\ \bibnamefont{Rossnagel}},\ }%
  \bibfield{journal}{%
  \bibinfo {journal} {Nano Lett.}\ }%
  \textbf{\bibinfo {volume} {9}},\ \bibinfo {pages} {2044} (\bibinfo {year}
  {2009})%
  \bibAnnoteFile{NoStop}{nam09}%
\bibitem{taniguchi09}%
  \BibitemOpen
  \bibfield{author}{%
  \bibinfo {author} {\bibfnamefont{M.}~\bibnamefont{Taniguchi}}, \bibinfo
  {author} {\bibfnamefont{M.}~\bibnamefont{Tsutsui}}, \bibinfo {author}
  {\bibfnamefont{K.}~\bibnamefont{Yokota}},\ and\ \bibinfo {author}
  {\bibfnamefont{T.}~\bibnamefont{Kawai}},\ }%
  \bibfield{journal}{%
  \bibinfo {journal} {Appl. Phys. Lett.}\ }%
  \textbf{\bibinfo {volume} {95}},\ \bibinfo {pages} {123701} (\bibinfo {year}
  {2009})%
  \bibAnnoteFile{NoStop}{taniguchi09}%
\bibitem{ivanov11}%
  \BibitemOpen
  \bibfield{author}{%
  \bibinfo {author} {\bibfnamefont{A.~P.}\ \bibnamefont{Ivanov}}, \bibinfo
  {author} {\bibfnamefont{E.}~\bibnamefont{Instuli}}, \bibinfo {author}
  {\bibfnamefont{C.~M.}\ \bibnamefont{McGilvery}}, \bibinfo {author}
  {\bibfnamefont{G.}~\bibnamefont{Baldwin}}, \bibinfo {author}
  {\bibfnamefont{D.~W.}\ \bibnamefont{McComb}}, \bibinfo {author}
  {\bibfnamefont{T.}~\bibnamefont{Albrecht}},\ and\ \bibinfo {author}
  {\bibfnamefont{J.~B.}\ \bibnamefont{Edel}},\ }%
  \bibfield{journal}{%
  \bibinfo {journal} {Nano Lett.}\ }%
  \textbf{\bibinfo {volume} {11}},\ \bibinfo {pages} {279} (\bibinfo {year}
  {2011})%
  \bibAnnoteFile{NoStop}{ivanov11}%
\bibitem{liu10}%
  \BibitemOpen
  \bibfield{author}{%
  \bibinfo {author} {\bibfnamefont{H.}~\bibnamefont{Liu}}, \bibinfo {author}
  {\bibfnamefont{J.}~\bibnamefont{He}}, \bibinfo {author}
  {\bibfnamefont{J.}~\bibnamefont{Tang}}, \bibinfo {author}
  {\bibfnamefont{H.}~\bibnamefont{Liu}}, \bibinfo {author}
  {\bibfnamefont{P.}~\bibnamefont{Pang}}, \bibinfo {author}
  {\bibfnamefont{D.}~\bibnamefont{Cao}}, \bibinfo {author}
  {\bibfnamefont{P.}~\bibnamefont{Krstic}}, \bibinfo {author}
  {\bibfnamefont{S.}~\bibnamefont{Joseph}}, \bibinfo {author}
  {\bibfnamefont{S.}~\bibnamefont{Lindsay}},\ and\ \bibinfo {author}
  {\bibfnamefont{C.}~\bibnamefont{Nuckolls}},\ }%
  \bibfield{journal}{%
  \bibinfo {journal} {Science}\ }%
  \textbf{\bibinfo {volume} {327}},\ \bibinfo {pages} {64} (\bibinfo {year}
  {2010})%
  \bibAnnoteFile{NoStop}{liu10}%
\end{thebibliography}%
\end{document}